\documentclass[letterpaper, 10 pt, conference]{ieeeconf}  

\IEEEoverridecommandlockouts                              

\usepackage{graphicx}
\usepackage{mathcomp}
\usepackage{booktabs}
\usepackage{tablefootnote}
\usepackage{threeparttable}
\UseRawInputEncoding

\title{\LARGE \bf 
Neurophysiological Response Based on Auditory Sense for Brain Modulation Using Monaural Beat 
}

\author{Ha-Na Jo$^{1}$, Young-Seok Kweon$^{2}$, Gi-Hwan Shin$^{2}$, Heon-Gyu Kwak$^{1}$, and Seong-Whan Lee$^{1}$
\thanks{*This work was partly supported by Institute for Information $\&$ Communications Technology Promotion (IITP) grants funded by the Korea government(MSIT) (No. 2017-0-00451: Development of BCI based Brain and Cognitive Computing Technology for Recognizing User's Intentions using Deep Learning, No. 2019-0-00079: Artificial Intelligence Graduate School Program, Korea University, and No. 2021-0-02068: Artificial Intelligence Innovation Hub).}
\thanks{$^{1}$Department of Artificial Intelligence, Korea University, Seoul, Republic of Korea}%
\thanks{$^{2}$Department of Brain and Cognitive Engineering, Korea University, Seoul, Republic of Korea}%
}

\begin{document}

\maketitle

\thispagestyle{empty}
\pagestyle{empty}

\begin{abstract}

Brain modulation is a modification process of brain activity through external stimulations. However, which condition can induce the activation is still unclear. Therefore, we aimed to identify brain activation conditions using 40 Hz monaural beat (MB). Under this stimulation, auditory sense status which is determined by frequency and power range is the condition to consider. Hence, we designed five sessions to compare; no stimulation, audible (AB), inaudible in frequency, inaudible in power, and inaudible in frequency and power. Ten healthy participants underwent each stimulation session for ten minutes with electroencephalogram (EEG) recording. For analysis, we calculated the power spectral density (PSD) of EEG for each session and compared them in frequency, time, and five brain regions. As a result, we observed the prominent power peak at 40 Hz in only AB. The induced EEG amplitude increase started at one minute and increased until the end of the session. These results of AB had significant differences in frontal, central, temporal, parietal, and occipital regions compared to other stimulations. From the statistical analysis, the PSD of the right temporal region was significantly higher than the left. We figure out the role that the auditory sense is important to lead brain activation. These findings help to understand the neurophysiological principle and effects of auditory stimulation.


\end{abstract}


\section{INTRODUCTION} 

Brain modulation aims to induce brain activity using various stimulations. Transcranial electric and magnetic stimulation are one of the various methods of brain modulation \cite{electric, magnetic}. But these methods have adverse effects that contain mild tingling sensation, fatigue, and headache \cite{safety}. 

Unlike these direct transcranial-based stimulations, binaural beat (BB) and monaural beat (MB) are indirect auditory-based stimulations \cite{introBB, embc1}. These beats consist of lower and upper frequencies called the carrier and offset \cite{component}. The BB is a stereo sound using sine waves of neighboring frequencies into each ear, whereas the MB is a mono sound that combines into one channel \cite{component, embc2}. These auditory-based stimulations can induce the activation of brain signals based on the differences in the carrier and offset frequencies, which can be monitored by the frequency-following response (FFR) \cite{ffr1}. Moreover, these do not need any direct device to the scalp, just earphones or speakers.

Auditory stimulation like BB and MB are recognized by the auditory sense which is a simple automatic reaction to sounds by an auditory organ. It is different than perception which needs more complex and requires effort for meaningful interpretations of experiences \cite{sense}. When people are subjected to stimulation by BB and MB, they can sense the integrated sound of carrier and offset frequency, but they can not perceive that the sound is a combination of two different frequencies \cite{percept}. Also, whether or not auditory sensation is possible in these stimulations is determined by the frequency and power range \cite{audiogram}. If the frequency is higher than the hearing threshold or the power is smaller than the threshold, people can not sense the BB and MB. Most studies observed the effect of auditory stimulations which could sense \cite{alphaMB, gammaBB}. However, these could not consider inaudible stimulation that could not sense. 

In this study, we aimed to find the activation condition of auditory stimulation in terms of auditory sense. In detail, MB was used as auditory stimulation to confirm the changes in EEG amplitude clearly. Moreover, we designed the experiment to compare the outcomes in participants after subjecting them to both audible and inaudible conditions, considering both frequency and power range. In consequence, we could identify that the auditory sensation has an essential role in inducing brain activity.

\begin{figure*}[thpb]
  \centering
  \includegraphics[width=\textwidth]{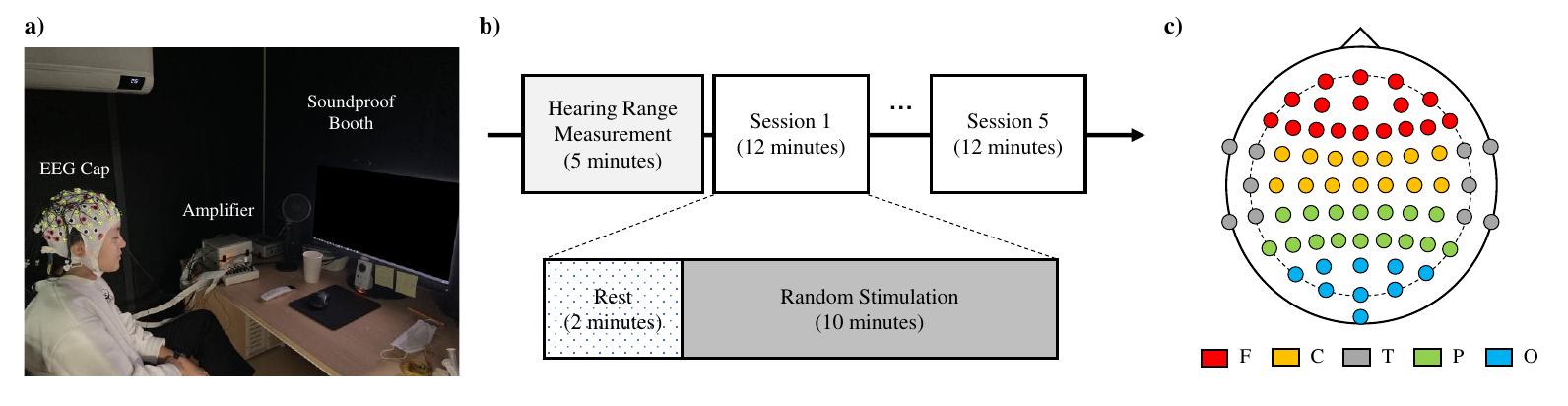}
  \caption{Experimental setting. a) Experimental environment. b) Experimental procedure. Random stimulation consists of no stimulation, audible monaural beat, inaudible monaural beat in frequency, inaudible monaural beat in power, and inaudible monaural beat in frequency and power. c) Segmentation of brain regions. The brain region is divided into five groups: F=frontal, C=central, T=temporal, P=parietal, O=occipital.}
  \label{figure2}
\end{figure*}

\section{Methods}

\subsection{Participants and Experimental Procedure}
Ten participants were included in this study (five males and five females, mean ages 26.1$\pm$3.4 years). None of the participants had a medical history of claustrophobia or hearing loss. This study was approved by the Institutional Review Board at Korea University (KUIRB-2022-0222-01), and each participant provided written informed consent before the start of experiment.

The participants visited the soundproof booth and prepared for electroencephalogram (EEG) acquisition (Fig. \ref{figure2}-a). Individual hearing thresholds were measured (mean hearing downer threshold 9.5$\pm$3.3 dB) and five sessions with different stimulation conditions were performed randomly. Each session consisted of two minutes of rest and ten minutes of stimulation. It is known as the optimal time to activate the brain \cite{component, 10min}. During the experiments, the participants kept their eyes closed (Fig. \ref{figure2}-b). 

\subsection{Audible and Inaudible Monaural Beat} 
The MB, which has a greater amplitude and a broader frequency range than the BB, was used in the experiment \cite{introBB}. The targeted frequency was set to 40 Hz which has a more significant auditory potential response than different frequencies \cite{40hz}. We presented the no stimulation and four conditions of the 40 Hz MB: no stimulation (NS), audible (AB), inaudible in frequency (IB-f), inaudible in power (IB-p), and inaudible in frequency and power (IB). For the AB, we set the carrier tone to 400 Hz which performs better than other frequencies and the offset tone of stimulation to 440 Hz \cite{400hz}. The sound intensity was set to 40 dB to activate the auditory cortex \cite{40db}. For IB-f, the carrier frequency was 18,000 Hz and the offset was 18,040 Hz. The power was the same as that used for AB. For IB-p, we set the power to 5 dB based on the fact that we tested a hearing threshold. The frequency was the same as the audible condition. For IB, we combined the frequency setting of IB-f and the power setting of IB-p. All stimulations were created using the audio generator program Gnaural 1.0.20110606.

\subsection{EEG Recording and Analysis}
We recorded the EEG signal at a 1,000 Hz sampling rate using an amplifier (BrainAmp; Brian Products, Germany). The 64 channels, using Ag/AgCI electrodes, were placed according to the 10-20 international system. EEG pre-processing was performed with the EEGLAB toolbox for MATLAB following \cite{eeglab, brainregion} and consisted of down-sampled to 250 Hz, band pass filtered between 0.5 Hz to 50 Hz, and channel interpolation using the Gaussian distribution’s kurtosis. 

The recorded EEG signals were segmented into one minute intervals to observe changes over time. Additionally, we divided brain regions into five groups as follows: frontal, central, temporal, parietal, and occipital regions (Fig. \ref{figure2}-c) \cite{brainregion}. To identify brain symmetry, each region was divided into left and right, excluding the center electrodes (FPz, AFz, Fz, FCz, Cz, CPz, Pz, POz, Oz, and Iz). 

For neurophysiological analysis, the power spectral density (PSD) was estimated using a fast Fourier transform. The power ratio ($R_f$) was calculated to identify the trend changes in PSD for each frequency and defined as
\begin{equation} 
	R_{f}=\frac{2\times PSD_{f}}{PSD_{f-1}+PSD_{f+1}},
\end{equation}
where the frequency ($f$) for the analysis was set from the 1 Hz to 50 Hz range. In the case of a constant tendency, $R_f$ converges to one, otherwise, it diverges to infinity.

All data were analyzed using one-way analysis of variance (ANOVA) method. In the case of significance, a post-hoc analysis was performed using \textit{t}-test. The significance level was set at \textit{p}$<$0.05.

\section{Results}

\subsection{Frequency-following Response by Stimulation}  
\begin{figure}[thpb]
  \centering
  \includegraphics[width=240pt]{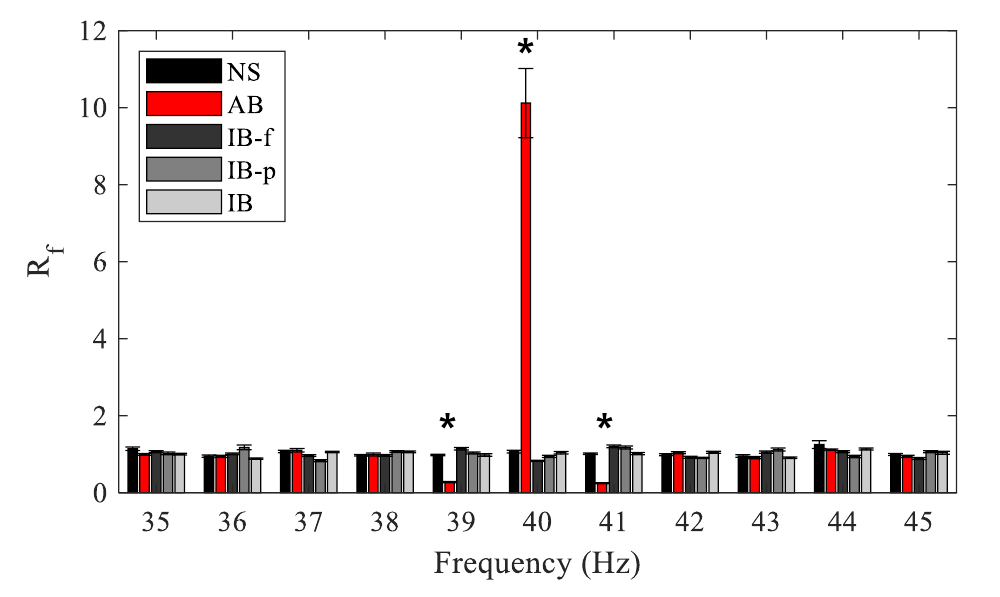}
  \caption{Power ratio nearby targeted frequency (40 Hz). $R_f$=Power ratio, NS=no stimulation, AB=audible monaural beat, IB-f=inaudible monaural beat in frequency, IB-p=inaudible monaural beat in power, IB=inaudible monaural beat in frequency and power. The error bars indicate standard errors. * represents the statistical significance (\textit{p}$<$0.05).}
  \label{figure3}
\end{figure}

Fig. \ref{figure3} shows the power ratio near the targeted frequency (40 Hz) to identify the EEG amplitude changes caused by the stimulation. We observed a considerably increased response at 40 Hz in only one of the sessions, AB (NS: $R_{40}$=1.07, AB: $R_{40}$=10.11, IB-f: $R_{40}$=0.83, IB-p: $R_{40}$=0.94, IB: $R_{40}$=1.04) with significance (NS: \textit{p}$=$0.01, IB-f: \textit{p}$=$0.01, IB-p: \textit{p}$=$0.01, IB: \textit{p}$=$0.01). In comparison, a significantly smaller change in response was observed at 39 Hz and 41 Hz for AB (\textit{p}$<$0.001, each). In other stimulations, including NS and inaudible conditions, there were no negative or positive peaks at 40 Hz (NS vs. IB-f: \textit{p}=0.07, NS vs. IB-p: \textit{p}=0.42, NS vs. IB: \textit{p}=0.83, IB-f vs. IB-p: \textit{p}=0.29, IB-f vs. IB: \textit{p}=0.15, IB-p vs. IB: \textit{p}=0.35). We also explored the constant tendency in other frequencies excluding 40 Hz in all sessions and none of the sessions showed any differences.

\begin{figure}[t]
  \centering
  \includegraphics[width=240pt]{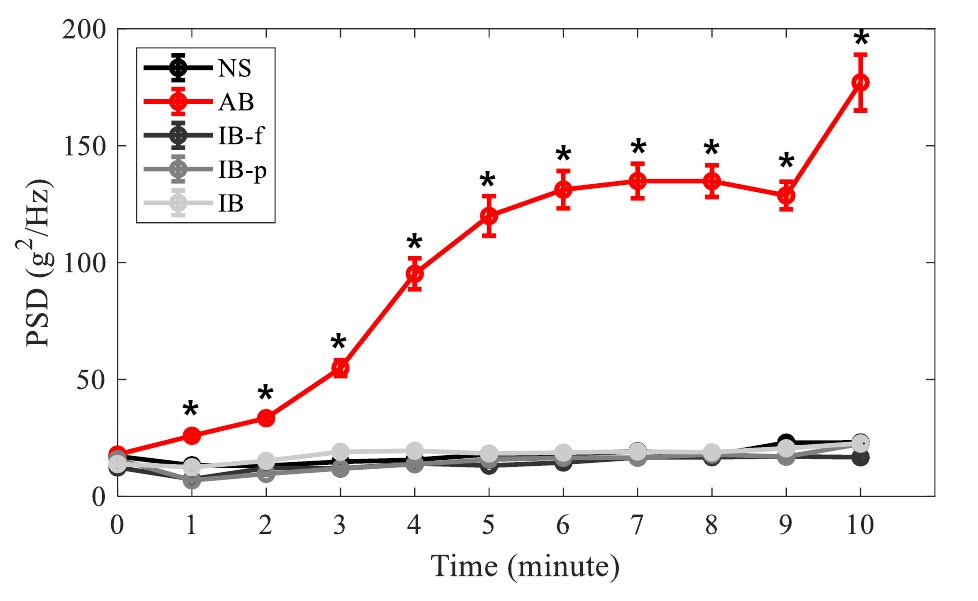}
  \caption{Power spectral density (PSD) from 40 Hz for ten minutes stimulation phase in each stimulation. NS=no stimulation, AB=audible monaural beat, IB-f=inaudible monaural beat in frequency, IB-p=inaudible monaural beat in power, IB=inaudible monaural beat in frequency and power. The error bars indicate standard errors. * represents the statistical significance (\textit{p}$<$0.05).}
  \label{figure4}
\end{figure}

\begin{figure*}[thpb]
  \centering
  \includegraphics[width=\textwidth]{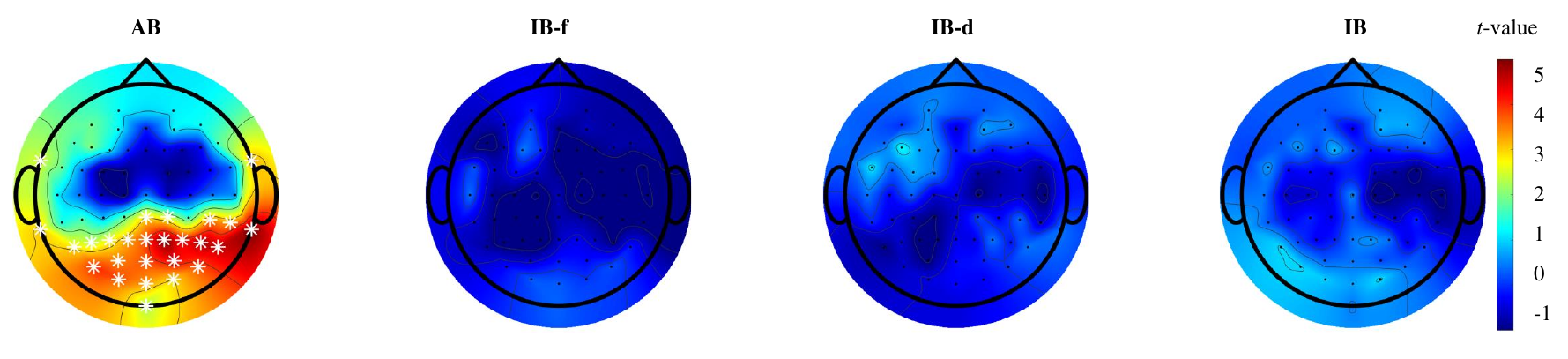}
  \caption{Topographical distribution of the \textit{t}-value of the comparison from 40 Hz power spectral density between no stimulation (NS) and other sessions: audible monaural beat (AB), inaudible monaural beat in frequency (IB-f), inaudible monaural beat in power (IB-p), inaudible monaural beat in frequency and power (IB). * represents the statistical significance (\textit{p}$<$0.05).}
  \label{figure5}
\end{figure*}

Fig. \ref{figure4} shows the results of time-related spectral analysis at 40 Hz of the segmented stimulation phase. Baseline values (before stimulation) were essentially similar among the five experimental groups. However, in the stimulation phase, we observed a statistically significant power peak during AB compared with other stimulations (\textit{p}$<$0.001, each time segment). During the first minute, activation of the targeted frequency in AB group was significantly higher than in IB-f and IB-p, while it was only slightly higher compared to other two groups (NS: \textit{p}$=$0.09, IB-f: \textit{p}$<$0.01, IB-p: \textit{p}$<$0.01, IB: \textit{p}$=$0.06), however, it increased significantly compared to other stimulations until the end of the phase (\textit{p}$<$0.05, each). No difference was observed between the NS and inaudible conditions.

Moreover, we compared the PSD changes at 40 Hz. Fig. \ref{figure5} shows the topology of the \textit{t}-value of the comparison between NS and the other stimulations. We observed significant activation of some of channels in the parietal, temporal, and occipital regions in AB. However, there were no differences between the other sessions and NS. 

\subsection{Asymmetric activation by stimulation} 
We compared the PSD changes at 40 Hz in all brain regions (Table. \ref{table1}). There was no statistically significant difference, for any regions, between the other four groups, excluding AB. However, significantly larger PSD values were obtained for AB compared to the other stimulation groups. In detail, for the frontal region, there was a difference only between AB and IB-f, not others (NS: \textit{p}$=$0.05, IB-f: \textit{p}$=$0.02, IB-p: \textit{p}$=$0.1, IB: \textit{p}$=$0.09). Additionally, for the central region, AB group had more pronounced activation than the others four groups (NS: \textit{p}$=$0.01, IB-f: \textit{p}$=$0.003, IB-p: \textit{p}$=$0.01, IB: \textit{p}$=$0.01). The inducing effect of AB in the temporal region was also significant (NS: \textit{p}$=$0.01, IB-f: \textit{p}$=$0.002, IB-p: \textit{p}$=$0.004, IB: \textit{p}$=$0.01). In the parietal region, differences between AB and other groups were statistically significant (NS: \textit{p}$=$0.002, IB-f: \textit{p}$=$0.001, IB-p: \textit{p}$=$0.001, IB: \textit{p}$=$0.002). Similar results were obtained for the occipital region (NS: \textit{p}$=$0.001, IB-f: \textit{p}$<$0.001, IB-p: \textit{p}$=$0.002, IB: \textit{p}$=$0.002).

We also carried out a brain symmetry comparison between left and right brain regions (Table. \ref{table2}). The PSD was only calculated for AB group as only this group showed targeted brain activity after stimulation. The brain response was symmetrical in the frontal, central, parietal, and occipital regions. However, in the temporal region, the right brain was significantly more activated compared to the left brain. 


\begin{table}[]
\centering
\caption{Comparison of power spectral density from 40 Hz between stimulations.}
\label{table1}
\resizebox{\columnwidth}{!}{%
\Large
\setlength{\tabcolsep}{8pt} 

\begin{tabular}{c|ccccc|c}
\hline
\textbf{Region}    & \textbf{NS} & \textbf{AB}     & \textbf{IB-f} & \textbf{IB-p} & \textbf{IB} & \textbf{\textit{p}-value}           \\ \hline
Frontal   & 25.93         & \textbf{55.88}  & 20.56         & 28.31         & 27.13       & \textbf{0.01}              \\
Central   & 114.76        & \textbf{526.26} & 42.82         & 70.01         & 96.18       & \textbf{\textless{}0.001} \\
Temporal  & 14.34         & \textbf{34.21}  & 9.86          & 10.80         & 12.22       & \textbf{\textless{}0.001} \\
Parietal  & 30.52         & \textbf{200.47} & 24.73         & 27.82         & 36.86       & \textbf{\textless{}0.001} \\
Occipital & 95.40         & \textbf{620.33} & 104.70        & 91.34         & 118.79      & \textbf{\textless{}0.001} \\ \hline
\end{tabular}%
}
\begin{tablenotes}[flushleft]
\item AB=audible monaural beat, IB-f=inaudible monaural beat in frequency, IB-p=inaudible monaural beat in power, IB=inaudible monaural beat in frequency and power. We marked the statistical significance in bold.
\end{tablenotes}
\end{table}

\begin{table}[]
\centering
\caption{Comparison of power spectral density from 40 Hz between right and left brain in audible monaural beat.}
\label{table2}
\resizebox{\columnwidth}{!}{%
\setlength{\tabcolsep}{8pt} 
\tiny
\begin{tabular}{c|cc|c} 
\hline
\textbf{Region} & \textbf{Right}  & \textbf{Left}   & \textbf{\textit{p}-value} \\ \hline
Frontal        & 83.53           & 64.36           & 0.19             \\
Central         & 61.84           & 39.46           & 0.16             \\
Temporal        & \textbf{670.30} & \textbf{432.50} & \textbf{0.02}    \\
Parietal        & 278.74          & 180.60          & 0.07             \\
Occipital       & 686.55          & 511.88          & 0.13             \\ \hline
\end{tabular}%
}
\begin{tablenotes}[flushleft]
\item We marked the statistical significance in bold.
\end{tablenotes}
\end{table}


\section{Discussion and Conclusion} 
This study investigated the relationship between auditory senses and brain activation, especially FFR. We compared the audible and inaudible conditions and observed that only AB could significantly induce brain activity at 40 Hz. Additionally, its effects increased over stimulation time and were slightly more extensive in the right brain. 

Our results showed that AB caused FFR in all brain regions, and the right temporal region was more activated than other regions throughout the stimulation time. This result is in agreement with the results of recent studies that used BB or MB in the human audible range \cite{alphaMB, gammaBB, 40db}. Exceptionally, in the power ratio analysis in this study, there were two additional significantly different points. However, this phenomenon occurred because a prominent power peak at 40 Hz affected the ratio formula of the two sides and, not because of any change in brain activity. A comparison of PSD between right and left regions was also in accordance with the results of previous studies \cite{ffr1, ffr2}. These studies used auditory click stimulation, such as speech or piano for fewer than 30 seconds and observed activation from the right auditory cortex. Therefore, only AB caused targeted FFR in all brain regions, and a right-asymmetric effect was observed for the temporal region. 

Unlike AB, we could not find an increasing or decreasing tendency in the amplitude during exposure to inaudible MBs. According to its essential characteristics, the MB is known to have a more comprehensive carrier frequency range than the BB and a recent study compared its effectiveness up to 3,000 Hz \cite{introBB, 400hz}. We explored the utilization of MB over 3,000 Hz in extension. In terms of inaudible powers, to the best of our knowledge, no previous study has carried out such a comparison, and ours is the first study of this kind. Since inaudible MBs were inaudible and didn't involve any sound processing in the brain, our results provided evidence that FFR is a by-product of sound processing. 

In conclusion, we investigated the role of auditory senses in brain activation in response to MB. The auditory sense in terms of frequency and power is essential to lead brain activation. Our study could help to understand the principle of FFR by auditory stimulation. In addition, these findings should be considered important in various fields of brain modulation using MB, such as education to improve concentration, treatment, sleep technology to induce sleep, and brain-computer interfaces when using sounds.

\bibliographystyle{IEEEtran-mod}
\bibliography{reference}

\begin{thebibliography}{10}
\providecommand{\url}[1]{#1}
\csname url@samestyle\endcsname
\providecommand{\newblock}{\relax}
\providecommand{\bibinfo}[2]{#2}
\providecommand{\BIBentrySTDinterwordspacing}{\spaceskip=0pt\relax}
\providecommand{\BIBentryALTinterwordstretchfactor}{4}
\providecommand{\BIBentryALTinterwordspacing}{\spaceskip=\fontdimen2\font plus
\BIBentryALTinterwordstretchfactor\fontdimen3\font minus
  \fontdimen4\font\relax}
\providecommand{\BIBforeignlanguage}[2]{{%
\expandafter\ifx\csname l@#1\endcsname\relax
\typeout{** WARNING: IEEEtran.bst: No hyphenation pattern has been}%
\typeout{** loaded for the language `#1'. Using the pattern for}%
\typeout{** the default language instead.}%
\else
\language=\csname l@#1\endcsname
\fi
#2}}
\providecommand{\BIBdecl}{\relax}
\BIBdecl

\bibitem{electric}
P.~Minhas, M.~Bikson, A.~J. Woods, A.~R. Rosen, and S.~K. Kessler,
  ``{Transcranial Direct Current Stimulation in Pediatric Brain: a
  Computational Modeling Study},'' in \emph{Proc. International Conference of
  the IEEE Engineering in Medicine and Biology Society (EMBC)}, San Diego,
  2012, pp. 859--862.

\bibitem{magnetic}
M.~Massimini, F.~Ferrarelli, S.~K. Esser, B.~A. Riedner, R.~Huber, M.~Murphy,
  M.~J. Peterson, and G.~Tononi, ``{Triggering Sleep Slow Waves by Transcranial
  Magnetic Stimulation},'' \emph{Proc. Natl. Acad. Sci. U.S.A.}, vol. 104,
  no.~20, pp. 8496--8501, May 2007.

\bibitem{safety}
C.~Poreisz, K.~Boros, A.~Antal, and W.~Paulus, ``{Safety Aspects of
  Transcranial Direct Current Stimulation Concerning Healthy Subjects and
  Patients},'' \emph{Brain Res. Bull.}, vol.~72, no. 4-6, pp. 208--214, May
  2007.

\bibitem{introBB}
L.~Chaieb, E.~C. Wilpert, T.~P. Reber, and J.~Fell, ``{Auditory Beat
  Stimulation and its Effects on Cognition and Mood States},'' \emph{Front.
  Psychiatry}, vol.~6, p.~70, May 2015.

\bibitem{embc1}
T.~Yamsa-Ard and Y.~Wongsawat, ``{The Observation of Theta Wave Modulation on
  Brain Training by 5 Hz-Binaural Beat Stimulation in Seven Days},'' in
  \emph{Proc. 37th Annu. International Conference of the IEEE Engineering in
  Medicine and Biology Society (EMBC)}, Milan, 2015, pp. 6667--6670.

\bibitem{component}
N.~Jirakittayakorn and Y.~Wongsawat, ``{Brain Responses to a 6-Hz Binaural
  Beat: Effects on General Theta Rhythm and Frontal Midline Theta Activity},''
  \emph{Front. Neurosci.}, vol.~11, p. 365, Jun. 2017.

\bibitem{embc2}
N.~Jirakittayakorn and Y.~Wongsawat, ``{The Brain Responses to Different
  Frequencies of Binaural Beat Sounds on QEEG at Cortical Level},'' in
  \emph{Proc. 37th Annu. International Conference of the IEEE Engineering in
  Medicine and Biology Society (EMBC)}, Milan, 2015, pp. 4687--4691.

\bibitem{ffr1}
E.~B. Coffey, S.~C. Herholz, A.~M. Chepesiuk, S.~Baillet, and R.~J. Zatorre,
  ``{Cortical Contributions to the Auditory Frequency-following Response
  Revealed by MEG},'' \emph{Nat. Commun.}, vol.~7, no.~1, pp. 1--11, Mar. 2016.

\bibitem{sense}
G.~Mather, \emph{{Foundations of Sensation and Perception}}.\hskip 1em plus
  0.5em minus 0.4em\relax London: Psychology Press, 2016, p. 3.

\bibitem{percept}
G.~Oster, ``{Auditory Beats in the Brain},'' \emph{Sci. Am.}, vol. 229, no.~4,
  pp. 94--103, Oct. 1973.

\bibitem{audiogram}
L.~L. Jackson, R.~S. Heffner, and H.~E. Heffner, ``{Free-field Audiogram of the
  Japanese Macaque (Macaca Fuscata)},'' \emph{J. Acoust. Soc. Am.}, vol. 106,
  no.~5, pp. 3017--3023, Oct. 1999.

\bibitem{alphaMB}
L.~Chaieb, E.~C. Wilpert, C.~Hoppe, N.~Axmacher, and J.~Fell, ``{The Impact of
  Monaural Beat Stimulation on Anxiety and Cognition},'' \emph{Front. Hum.
  Neurosci.}, vol.~11, p. 251, May 2017.

\bibitem{gammaBB}
N.~Jirakittayakorn and Y.~Wongsawat, ``{Brain Responses to 40-Hz Binaural Beat
  and Effects on Emotion and Memory},'' \emph{Int. J. Psychophysiol.}, vol.
  120, pp. 96--107, Oct. 2017.

\bibitem{10min}
P.~Goodin, J.~Ciorciari, K.~Baker, A.-M. Carrey, M.~Harper, and J.~Kaufman,
  ``{A High-density EEG Investigation into Steady State Binaural Beat
  Stimulation},'' \emph{PloS one}, vol.~7, no.~4, p. e34789, Apr. 2012.

\bibitem{40hz}
M.~A. Pastor, J.~Artieda, J.~Arbizu, J.~M. Marti-Climent, I.~Pe{\~n}uelas, and
  J.~C. Masdeu, ``{Activation of Human Cerebral and Cerebellar Cortex by
  Auditory Stimulation at 40 Hz},'' \emph{J. Neurosci.}, vol.~22, no.~23, pp.
  10\,501--10\,506, Dec. 2002.

\bibitem{400hz}
D.~W. Schwarz and P.~Taylor, ``{Human Auditory Steady State Responses to
  Binaural and Monaural Beats},'' \emph{Clin. Neurophysiol.}, vol. 116, no.~3,
  pp. 658--668, Mar. 2005.

\bibitem{40db}
M.~R{\"o}hl and S.~Uppenkamp, ``{Neural Coding of Sound Intensity and Loudness
  in the Human Auditory System},'' \emph{J. Assoc. Res. Otolaryngol.}, vol.~13,
  pp. 369--379, Feb. 2012.

\bibitem{eeglab}
A.~Delorme and S.~Makeig, ``{EEGLAB: an Open Source Toolbox for Analysis of
  Single-trial EEG Dynamics Including Independent Component Analysis},''
  \emph{J. Neurosci. Methods}, vol. 134, no.~1, pp. 9--21, Mar. 2004.

\bibitem{brainregion}
G.-H. Shin, M.~Lee, and S.-W. Lee, ``{Assessment of Unconsciousness for Memory
  Consolidation using EEG Signals},'' in \emph{Proc. IEEE International
  Conference on Systems, Man, and Cybernetics (SMC)}, Toronto, 2020, pp.
  513--519.

\bibitem{ffr2}
E.~B. Coffey, G.~Musacchia, and R.~J. Zatorre, ``{Cortical Correlates of the
  Auditory Frequency-following and Onset Responses: EEG and fMRI Evidence},''
  \emph{J. Neurosci.}, vol.~37, no.~4, pp. 830--838, Jan. 2017.

\end{thebibliography}

\end{document}